\documentclass[11pt]{article}
\usepackage{hyperref}

\usepackage{tikz-cd}
\usepackage{xcolor}
\usepackage{amsmath,amsthm,amsfonts,amssymb,amsopn,amscd,bm,cancel} 
\usepackage{color}

\def\qed{{\unskip\nobreak\hfil\penalty50
\hskip2em\hbox{}\nobreak\hfil$\square$
\parfillskip=0pt \finalhyphendemerits=0\par}\medskip}
\def\proof{\trivlist \item[\hskip \labelsep{\bf Proof.\ }]}

\def\eproof{\null\hfill\qed\endtrivlist\noindent}

\def\tilde{\widetilde}

\def\a{\alpha}

\def\l{\lambda}

\def\Om{\Omega}

\def\A{{\cal A}}

\def\R{{\cal R}}

\def\H{{\cal H}}

\def\S{{\cal S}}

\def\f{{\varphi}}

\def\l{{\lambda}}
\def\x{{{h}}}

\def\PSL{{{\rm PSL}(2,\mathbb R)}}

\def\S2{S^{1(2)}}

\def\RR{\mathbb R}

\newtheorem{theorem}{Theorem}[section]
\newtheorem{lemma}[theorem]{Lemma}

\newtheorem{corollary}[theorem]{Corollary}

\newtheorem{proposition}[theorem]{Proposition}

\theoremstyle{remark} 

\newcommand{\ben}{\begin{equation}}
\newcommand{\een}{\end{equation}}

\newcommand{\bthm}{\begin{theorem}}
\newcommand{\ethm}{\end{theorem}}

\newcommand{\bprop}{\begin{proposition}}
\newcommand{\eprop}{\end{proposition}}
\newcommand{\bcor}{\begin{corollary}}
\newcommand{\ecor}{\end{corollary}}
\newcommand{\blem}{\begin{lemma}}
\newcommand{\elem}{\end{lemma}}

\def\PSL{PSU(1,1)}

\def\CC{{\mathbb C}}

\def\SL2{{{\rm SL}(2,\R)}}

\def\PSL2{{{\rm PSL}(2,\Reali)}}

\def\U1{{{\rm V}(1)}}
\def\SU2{{{\rm SV}(2)}}

\def\SU{{{\rm SU}}}

\def\A{{\mathcal A}}

\def\H{{\mathcal H}}

\def\O{{\mathcal O}}

\def\S{{\mathcal S}}

\def\W{{\mathcal W}}

\def\RR{{\mathbb R}}

\def\a{\alpha}

\def\l{\lambda}       \def\L{\Lambda}

\def\x{\xi}

\def\f{\varphi}

\def\o{\omega}        

\def\S{{S(\RR^d)}}
\def\ov{\overline}

\title{\Huge{A Bekenstein-type bound in QFT}
} 

\author{ {\sc Roberto  Longo}\\
Dipartimento di Matematica,
Tor Vergata Universit\`a di Roma\\
Via della Ricerca Scientifica, 1, I-00133 Roma, Italy
}

\date{}
\begin{document}

\maketitle

\begin{abstract}
Let $B$ be a spacetime region of width $2R >0$, and $\f$  a vector state localized in $B$. We show that the vacuum relative entropy of $\f$, on the local von Neumann algebra of $B$, is bounded by $2\pi R$-times the energy of  the state $\f$ in $B$. 

This bound is model-independent and rigorous; it follows solely from first principles in the framework of translation covariant, local Quantum Field Theory on the Minkowski spacetime. 
\end{abstract}

\newpage

\section{Introduction}
Bekenstein argued in the early 80s  \cite{B81} that an 
upper limit  is to exist for the entropy/energy ratio
of a bounded system 
\ben\label{Bek}
S \leq 2\pi \textit{\small R}\,  E \, ;
\een
here $S$, $E$ are the entropy and the energy contained in the system of effective radius $R$.

This limit was initially inferred from the generalised second low of thermodynamics, in particular from the Bekenstein-Hawking geometric interpretation of black hole entropy as proportional to the area of the black hole horizon, and proposed to be universal. 

In 2008, Casini \cite{Ca08} gave a precise formulation of this inequality by subtracting the vacuum entropy and energy contribution to eliminate divergencies
and reducing the bound \eqref{Bek} to the positivity of the relative entropy between a state $\f$ and the vacuum state $\o$ on the algebra of $B$
\ben\label{SK}
S(\f|\!| \o) = \langle K\rangle_\f - \delta S \geq 0\, ,
\een
where $\delta S$ is the vacuum-subtracted von Neumann entropy of $\f$, and $\langle K\rangle_\f$ is the modular energy, namely the expectation value on $\f$ of the Tomita-Takesaki modular Hamiltonian. The use of the modular Hamiltonian is indeed one of the key points here. 

An identity as \eqref{SK} had rigorously been obtained in \cite{L97} concerning the incremental black hole entropy, and more recently was spelled out in \cite{LX18} in relation to Casini's work, and to the bound \cite{L18}. 

The physical interpretation of the Bekenstein limit is still a matter of discussion, particularly concerning the meaning of the black hole entropy and its information content, see \cite{HW}. A rigorous, model-independent, complete derivation of the bound \eqref{Bek} is not available yet, and local Quantum Field Theory \cite{H} offers a natural framework to this end. 

The main purpose of this paper is to provide a universal, rigorous bound on the relative entropy/energy ratio in a given region $B$, which is derived from first principles and is complete concerning the mathematical aspects. 

Suppose $B_0$ is a space region of width $2R>0$; so, $B_0$ lies within two parallel planes with distance $2R$; for example, $B_0$ could be a ball of radius $R$. For every vector state $\xi$, that is localized in $B$, we shall derive the bound
\ben\label{bound}
S(\f|\!| \o) \leq 2\pi \textit{\small R}\,  \langle P\rangle_\xi \, ,
\een
with $S(\f|\!| \o)$ the vacuum relative entropy of the state $\f = \langle\cdot\rangle_\xi$ given by $\xi$ on the observable von Neumann algebra $\A(B)$ associated with the causal envelop $B$ of $B_0$ in the Minkowski spacetime, and
$P$ the usual Hamiltonian. $\xi$ localized in $B$ means that $\langle\cdot\rangle_\xi$  agrees with the vacuum state on $\A(B')$, with $B'$ the causal complement of $B$. The inequality \eqref{bound} readily implies the universal bound
\[
S(\f|\!| \o) \leq 2\pi \textit{\small R}\, E(\f;B)\, ,
\]
where $E(\f;B)$ is the energy of $\f$ in $B$, which is defined by minimizing $\langle P\rangle_\eta$ over all vector states $\eta$ such 
$\langle \cdot\rangle_\eta = \langle \cdot\rangle_\xi$ on $\A(B')$. One may note some similarity with forms of the Bekenstein bound considered in \cite{P08} and \cite{BC13}. 

We assume that translation covariance with positive energy holds, but Poincaré covariance is not needed for our result. The locality property is assumed to hold in the maximal sense for wedge regions $W$ (causal envelop of spatial half-spaces), namely $\A(W') = \A(W)'$; yet wedge duality may be relaxed by assuming a stronger form of localization, i.e. $\langle \cdot\rangle_\xi$ restricts to the vacuum state on the commutant $\A(B)'$ of $\A(B)$.
Note that the region $B_0$ may be unbounded; it has to be bounded only in one direction; the use of the width scale in the Bekenstein inequality appeared in \cite{B03}.  

Now, the interpretation of our bound \eqref{bound} is at least two-fold. On the one hand, it gives a local bound on entropy/energy, so it
represents a form of the Bekenstein limit; on the other hand, it sets an upper bound to the relative entropy, while Casini's argument relies on a lower bound for the relative entropy, that is its non-negativity \eqref{SK}; so, in a sense, our inequality is complementary to the Bekenstein limit. 

Further comments, and a brief outlook from our work, are contained in the last section. 

\section{Preliminaries}
In this section, we provide the background material to derive our bound. 

\subsection{Order for unbounded operators}
Let $A$ be a positive, selfadjoint operator on a Hilbert space $\H$; we denote by $D(A)$ the domain of $A$.   If $\xi\in\H$, we shall always write 
\[
(\xi, A\xi) = ||A^{1/2}\xi ||^2\  \ {\rm if}\  \xi\in D(A^{1/2}), \ \ {\rm and}\ (\xi, A\xi) = \infty \ {\rm otherwise}
\]
({\it definition of $(\xi, A\xi)$ in the quadratic form sense}). 

Let now $B$ be an arbitrary selfadjoint operator on $\H$ and set $B_\pm = \pm E_\pm B$, where $E_\mp$ are the spectral projections of $B$ relative to $(0,-\infty]$ and $(0,\infty)$; thus $B_\pm$ are positive selfadjoint operators and $B = B_+ - B_-$. 

If $\xi\in \H$, we write
\[
(\xi, B\xi) < \infty \quad \text{if}\quad (\xi, B_+\xi) < \infty
\]
and
\[
(\xi, B\xi) > - \infty \quad \text{if}\quad (\xi, B_- \xi) < \infty
\]
if $(\xi, B\xi) < \infty$. 

We say that $(\xi, B\xi)$ is {\it well-defined} if either $(\xi, B \xi)< \infty$ or $(\xi, B \xi) > \infty$; in this case, we set
\[
(\xi, B \xi) = (\xi, B_+ \xi) - (\xi, B_- \xi) \, .
\]
If $(\xi, B\xi)$ is well-defined, then
\[
(\xi, B \xi) = \int_{-\infty}^\infty \l \, d\mu(\l)\, ,
\]
where $\mu$ is the spectral measure of $\xi$   with respect to  $B$.  

We shall use following elementary integral formula for the logarithm function:
\ben\label{log}
\log\l = \int_0^\infty f(\l,t) dt\, ,\quad \l>0\, .  
\een
with
\[
f(\l, t) = -\big( (t+1)^{-1} -\l (t+\l)^{-1}\big)/t \, ;
\]
note that 
\ben\label{+/-}
f(\l, t) < 0, \  \l <1,\quad  {\rm and} \quad f(\l, t) > 0, \ \l > 1\, .
\een
Now, if $A$ is positive selfadjoint and injective and $\xi\in \H$, we have
\[
(\xi , \log A\,\xi) =  \int_0^\infty \log \l \,d\mu(\l)
\]
provided $(\xi , \log A\,\xi)$ is well-defined, where $\mu$ is the spectral measure $A$ on $[0,\infty)$  with respect to  $\xi$.  
 Indeed 
\ben\label{welldef}
(\xi , \log A\,\xi) = (\xi , (\log A)_+\,\xi) -  (\xi , ( \log A)_- ,\xi) =  \int_0^1 \log \l \,d\mu(\l) +
 \int_1^{\infty} \log \l \,d\mu(\l) \, ,
\een
where one of the two integrals in the above formula is finite. 
\blem\label{propwd}
Let $A$ be positive selfadjoint and injective on $\H$, and $\xi\in\H$. 

If $(\xi , \log A\,\xi)$ is well-defined, then
\[
(\xi , \log A\,\xi) = \int_0^\infty (\xi,f(\l,A) \xi)d\l \, .
\]
If $(\xi , \log A\,\xi) < \infty$, then
\ben\label{At}
(\xi , \log A\,\xi) = \lim_{t\to 0^+} \frac{||A^{t/2}\xi||^2 - ||\xi||^2}{t}\, .
\een
\elem
\proof
We have
\ben\label{12}
(\xi , \log A\,\xi) =  \int_0^\infty \log \l \,d\mu(\l)  =  \int_0^1\log \l \,d\mu(\l)  +  \int_1^\infty \log \l \,d\mu(\l) 
\een
and
\[
 \int_0^1 \log \l \,d\mu(\l) =   \int_0^1 \int_0^\infty f(\l,t) d\l d\mu(t)
=  \int_0^\infty \int_0^1 f(\l,t) d\mu(t)d\l\, ,
\]
where the change of the order of the integrals is justified because $f(\l,t)\leq 0$ \eqref{+/-}. 
 
Similarly
\[
 \int_1^\infty \log \l \,d\mu(\l) =   \int_1^{\infty} \int_0^\infty f(\l,t) d\l d\mu(t)
=  \int_0^\infty \int_1^{\infty} f(\l,t) d\mu(t)d\l \, .
\]
Therefore,
\[
 \int_0^{\infty} \log \l \,d\mu(\l) =  
\int_0^\infty (\xi,f(\l,A) \xi)d\l 
\]
because one of the two integrals in  \eqref{12} is finite by assumption. 

Eq. \eqref{At} follows by Lebesgue monotone convergence theorem because, 
for a fixed  $0<\l<1$,
$\lim_{t\to 0^+} (\l^t - 1)/t = \log\l$
 decreasingly and $(\l^t - 1)/t \geq \log \l$, while, for  a fixed $\l > 1$, $\lim_{t\to 0^+} (\l^t - 1)/t = \log\l$ increasingly. 
\eproof
Let  $A,B$ be selfadjoint positive operators on $\H$. Recall that $A \leq B$ {\it means that}
\ben\label{AB}
{\rm Dom}(B^{1/2}) \subset {\rm Dom}(A^{1/2}) \ {\rm and}\  (\xi, A\xi) \leq  (\xi, B\xi) 
\ {\rm for\ all} 
\ \xi \in {\rm Dom}(B^{1/2})\, ;
\een
equivalently
\[
A \leq B \Longleftrightarrow (\xi, A\xi)  \leq (\xi, B\xi)\, , \quad \forall \xi\in \H\, ;
\]
morever,
\[
A \leq B\Longleftrightarrow (B + 1)^{-1} \leq (A + 1)^{-1}\, ,
\]
see \cite[Sect. 10.3]{Sch}. 
\smallskip

\emph{
If $A,B$ are any selfadjoint operators on $\H$, we shall write
$
A \leq B
$
if
$(\xi, A\xi)  \leq (\xi, B\xi)$ for all $\xi\in\H$ such that  both $(\xi, A \xi)$ and $(\xi, B\xi)$  are well-defined.}
\smallskip

If $A\leq B$, and the spectrum of $A$ is contained in $(0,\infty)$, then $\log A \leq \log B$ because the logarithm function is operator monotone; in case zero belongs to the continuous spectrum of $A$, more care is needed to apply this operator monotonicity. 
 \bprop\label{A<B}
Let  $A,B$ be selfadjoint, injective operators on $\H$ with $A\geq 0$ and $B\geq 0$. Then 
\[
A\leq B \implies \log A \leq \log B\, .
\] 
In particular, if $B\geq c$ with a constant $c>0$, then
$
(\xi, \log A\,\xi) \leq (\xi, \log B\,\xi)
$
for all $\xi\in\H$ such that $(\xi, \log A\,\xi)$ is well-defined. 
 \eprop
 \proof
 Let $\xi\in\H$ with $(\xi, \log A\,\xi)$ and $(\xi, \log B\,\xi)$
 well-defined. 
 By Lemma \ref{propwd}, we have 
\[
(\xi,\log A\,\xi) = \int_0^\infty (\xi, f(A, t) \xi) dt \leq
\int_0^\infty (\xi, f(B, t) \xi) dt =(\xi,\log B\,\xi)\, ,
\]
because $f(\l,t)$ is an increasing function of $\l$ at fixed $t$. 

The rest is clear because $B \geq c$ implies that $\log B$ is semi-bounded, so $(\xi, \log B\,\xi)$ is  well-defined for all $\x\in\H$. 
\eproof

\subsection{Standard subspaces}
Let $\H$ be a (complex) Hilbert space. A {\it standard subspace} $H\in\H$ is a closed, real linear subspace $H$ of $\H$ such that
$H_\CC = H + i H$ is dense in $\H$ and $H\cap iH = \{0\}$. We denote by $S_H$ the {\it Tomita anti-linear involution} $S_H : \xi + i\eta  \mapsto 
 \xi - i\eta $, $\xi,\eta \in  H$ and 
by $\Delta_H$ and $J_H$ the {\it modular operator} and {\it conjugation} of $H$ defined by the polar decomposition $S_H = J_H \Delta_H^{1/2}$. 
Clearly, $D(S_H) = D( \Delta_H^{1/2}) = H_\CC$; $\Delta_H$ is positive and non-singular and $J_H$ is an anti-unitary involution.  We have $J_H \Delta_H J_H = \Delta_H^{-1}$ and $S_{H'} = S^*_H$. 

The main relations are
\[
\Delta_H^{is} H =H\, ,\quad s\in\RR\, ,\qquad J_H H = H'\, .
\]
Here, $H'$ is the {\it symplectic complement} of $H$, the real orthogonal of $iH$. 

If $K\subset H$ is a closed, real linear subspace, then $K$ is a standard subspace of the closure $\bar  K_\CC$ of $K_\CC$, so $\Delta_K$ and $J_K$ are defined on $\bar K_\CC$. We may consider $\Delta_K$ and $J_K$ (and functions of $\Delta_K$) as operator on $\H$ by setting them equal to zero on the orthogonal of $\bar K_\CC$. 

We refer to \cite{L08}  for details, and to \cite{Str,Tak} for the general modular theory of von Neumann algebras. 
\blem\label{lemlogm}
Let $H\subset \H$ be a standard subspace. If
 $\xi\in H_\CC$, then $(\xi, \log\Delta_H \xi)$ is well-defined; indeed
 \[
 (\xi, \log \Delta_H \xi) \leq  ||S_H\xi||^2/e\, .
 \]
\elem
\proof
We have
\[
 (\xi, \log \Delta_H \xi) = (J_H \Delta_H^{1/2} S_H\xi,  \log \Delta_H J_H \Delta_H^{1/2} S_H\xi)
 = - (S_H\xi,  \Delta_H \log \Delta_H S_H\xi)  \leq   e^{-1} ||S_H\xi||^2
\]
because $-\l \log \l \leq 1/e$ for all $\l >0$. 
\eproof
We now recall the following known lemma, see \cite{BDL190, L08}. 
\blem\label{T} 
Let $H$ be a standard subspace of $\H$ and $K\subset H$ a closed, real linear subspace. 
We have $D(\Delta_K^{iz}) \subset D(\Delta_H^{iz})$ and $\Delta_H^{iz} \Delta_K^{-iz}$ is bounded if $-1/2 \leq \Im z \leq 0$. 
Setting
\[
T_{H,K}(z) = \ov{\Delta_H^{iz} \Delta_K^{-iz}}F\, , \quad -1/2 \leq \Im z \leq 0 \, ,
\]
(closure) ,
where  $F$ is the orthogonal projection onto $\bar K_\CC$,
the operator-valued map 
$z \mapsto T_{H,K}(z)$ is holomorphic in the strip 
$-1/2 < \Im z < 0$,  weakly continuous 
on $-1/2 \leq  \Im z \leq0$, and $||T_{H,K}(z)||\leq 1$. 
If $K$ is standard, 
\ben\label{TC}
T_{K',H'}(z) = T_{H,K}(-\bar z)^*\, .
\een
\elem
\proof
Let $\xi\in K_\CC  = D(\Delta_K^{1/2})$ and $\eta \in H'_\CC  = D(\Delta_H^{-1/2})$. The vector-valued functions $\Delta_K^{-iz}\xi$ and $\Delta_H^{iz}\eta  = \Delta_{H'}^{-iz}\eta $ are analytic in $-1/2 < \Im z < 0$, continuous on $-1/2 \leq \Im z \leq 0$; moreover 
$||\Delta_H^{iz}\eta ||\leq ||\eta ||$, because $||\Delta_H^{1/2}\eta || = ||J_K \Delta_H^{1/2}\eta || =||\eta ||$, and similarly for $\Delta_K^{-iz}\f$. 
It follows that $z\mapsto (\Delta_H^{-i\bar z}\eta , \Delta_K^{-iz}\xi)$ is also analytic in $-1/2 < \Im z < 0$, continuous on $-1/2 \leq \Im z \leq 0$, moreover
\[
|(\Delta_H^{-i\bar z}\eta , \Delta_K^{-iz}\xi)|_{\Im z = 0} \leq ||\eta ||\, ||\xi||
\]
and
\begin{multline*}
|(\Delta_H^{-i\bar z}\eta , \Delta_K^{-iz}\xi)|_{\Im z = -1/2} 
= |(\Delta_H^{1/2}\Delta_H^{-is}\eta ,\Delta_K^{-1/2} \Delta_K^{-is}\xi)|
= |(\Delta_H^{1/2}\Delta_H^{-is}\eta ,  S_H^2 \Delta_K^{-1/2} \Delta_K^{-is}\xi)|\\
= |(S_{H'}\Delta_{H'}^{-1/2}\Delta_H^{-is}\eta ,  S_K \Delta_K^{-1/2} \Delta_K^{-is}\xi)|
= |(J_{H'}\Delta_H^{-is}\eta ,  J_K \Delta_K^{-is}\xi)| \leq ||\eta ||\, ||\xi||
\end{multline*}
($z = s -  i/2$). Thus, by the three line theorem,
\[
|(\Delta_H^{-i\bar z}\eta , \Delta_K^{-iz}\xi)| \leq ||\eta ||\, ||\xi||\, ,\quad -1/2 \leq \Im z \leq 0\, .
\]
As $H_\CC$ is a core for $\Delta_H^{-i\bar z}$, it follows that $\Delta_K^{-iz}\xi$ belongs to the domain of the adjoint $\Delta_H^{iz}$ of 
$\Delta_H^{-i\bar z}$ and $||\Delta_H^{iz} \Delta_K^{-iz}\xi||\leq ||\xi||$, and this implies the lemma. 

The holomorphy of $T_{H,K}(z)$ on the interior of the strip then follows by the equivalence between weak and strong analyticity. 

The relation \eqref{TC} is clear. 
\eproof 
\bcor\label{HKD}
If $K\subset H$ is an inclusion of standard subspaces of $\H$, we have
\ben\label{DD}
\Delta^{\a}_H \leq \Delta^{\a}_K \, ,
\een
for all $0\leq \a \leq 1$, and
\ben\label{DD2}
-\log\Delta_K \leq - \log\Delta_H\, .
\een
\ecor
\proof
Let $0\leq \a \leq 1/2$. By Lemma \ref{T}, we have $||\Delta_H^\a\Delta_K^{-\a}\xi||\leq ||\xi||$ if $\xi\in D(\Delta_K^{-\a})$; therefore $||\Delta_H^\a\eta || \leq  ||\Delta_K^{\a}\eta ||$ for all $\eta  \in D(\Delta_K^{\a})$, so \eqref{DD} holds by \eqref{AB}. 

The inequality \eqref{DD2} then follows by Proposition \ref{A<B}. 
\eproof
The following is the standard subspace analogue of Borchers' theorem (in the von Neumann algebra setting) \cite{Bo92}.  Its proof, by the same arguments, was given in \cite{L08}.
\bthm\label{BH}
Let $\H$ be a Hilbert space and $U_\pm$  one-parameter unitary groups on $\H$ with positive generators  $ P_\pm \geq 0$. 

If $H\subset \H$ is a standard subspace and $U_\pm(t) H\subset H$ for all $\pm t \geq 0$, then
\[
\Delta_H^{is} U_\pm(t) \Delta_H^{- is}  = U_\pm(e^{\mp 2\pi s}t)\, ,\quad J_H U_\pm(t) J_H = U_\pm (-t)\, ,
\]
for all $t,s\in\RR$. 
\ethm
\subsection{Abstract entropy formulas}
In the following, if $A$ is a positive selfadjoint operator with support $p$ and $f$ a Borel function on $[0,+\infty)$, we write $f(A)$ for the operator $f(Ap)p$. 

Let $M$ be a von Neumann algebra $M$ on a Hilbert space $\H$, With $\o,\psi' $ normal states on $M$ and $M'$,  Connes' spatial derivative 
$\Delta(\o/\psi') = \frac{d\o}{d\psi'}$ is a positive selfadjoint operator on $\H$ with support $ee'$, where $e\in M$, $e'\in M'$ are the supports of $\o$ on $M$ and $\psi'$ on $M'$. 
For simplicity, in the following, we assume $\o$ to be faithful. 

Let $\f$ be a normal state on $M$ given by a vector $\xi\in\H$. Araki's relative entropy between $\f$ and $\o$ is defined by
\ben\label{Edef}
S(\f|\!| \o)  = - (\xi, \log \Delta(\o/\f')\xi )\, ,
\een
where $\f'$ is the faithful state $(\xi, \cdot \xi)$ on $M'$. See \cite{Ar76, Ar77} for definition in terms of the relative modular operator.  Note that
$(\xi, \log \Delta(\o/\f')\xi)$ is well-defined, indeed $S(\f|\!| \o)\geq 0$ or $S(\f|\!| \o) = +\infty$, see \cite{OP}. 

 Therefore, the following equivalent definition, essentially due to Uhlmann \cite{OP}, holds
\ben\label{UE}
S(\f|\!| \o) = -\lim_{t\to 0^+} \big((\xi, \Delta^t(\o/\f')\xi) - 1\big)/t
\een
by \eqref{At};
this avoids using the logarithmic function. 

$S(\f|\!| \o)$ does not depend on the vector representative $\xi$ for $\f$. Indeed, if $\x_1\in\H$ is another vector representative, there exists a partial isometry $w\in M$ with initial/final projection $e$, $e_1$  onto  $\ov{M'\xi}$, $\ov{M'\xi_1}$. Then, $\Delta(\o/\f_1') = w\Delta(\o/\f')w^*$, with $\f_1'$ the state $(\x_1, \cdot \xi_1)$ on $M'$, so
\[
(\xi_1, \log \Delta(\o/\f'_1)\xi_1) = (w\xi, (w\log \Delta(\o/\f')w^*)w\xi) = (\xi, w\log \Delta(\o/\f')\xi)\, .
\]
We assume for simplicity that the vector representative $\xi$ of $\f$ is cyclic for $M$, but this is not necessary. 
With $(D\f:D\o)_s \in M$ the Connes Radon-Nikodym cocyle
between $\f$ and $\o$, we have
\[
 \Delta(\o /\f')^{is} =  u_s \Delta(\f /\f')^{is} = u_s \Delta_\xi^{is}e\, ,\quad  s\in \RR\, ,
\]
where $u_s = (D\f:D\o)_s^*$; see \cite[Thm. 7.4]{Str} for the first equality, while the second equality follows
because 
\ben\label{dx}
\Delta(\xi /\f')^{is}  =   \Delta_\xi^{is}e\, , \quad s\in \RR\, ,
\een
with $e$ the support of $\f$ and $\Delta_\xi$ the modular operator of $eMe$  with respect to  $\xi$.  

So
\[
 =  (\xi, u_s\xi) = \f(u_s)\, ,\quad s\in\RR\, .
\]
As $\f(u_s)$ is the boundary value of a function analytic in the strip $-1 < \Im z < 0$ (see \cite[Thm. 3.1]{Str}), and the same is true for 
$(\xi, \Delta(\o /\f')^{is}\xi )$ by \eqref{dx} (KMS property), we then have 
\ben\label{CE}
(\xi,  \Delta(\o /\f')^{t}\xi ) =   \f(u_{-it})\, ,\quad 0 < t < 1\, ,
\een
where $\f(u_{-it}) = \underset{s\to -it}{\textnormal {anal.cont.\,}}  \f(u_s)$. 
\bprop\label{CE1}
Let $M$ be a von Neumann algebra, $\o$, $\f$ a normal states of $M$, with $\o$ faithful, as above. Then
\ben\label{Su}
S(\f|\!| \o) = -i\frac{d}{dt}\f(u_{-it}) \big|_{t=0} 
\een
with $u_t = (D\f:D\o)^*_t$. 

Here, $ -i\frac{d}{dt}\f(u_{-it}) \big|_{t=0} = \lim_{t\to 0+} \big(\f(u_{-it}) - 1\big)/it$; this limit exists, and it is either non-negative or infinite. 
\eprop
\proof
Apply eq. \eqref{CE} to formula \eqref{UE} and get the first equality in \eqref{Su}; the limit is non-negative or infinite as it is equal to $S(\f|\!| \o$). 
\eproof
Let $\Om\in\H$ be a cyclic and separating vector representative of $\o$, and  $v\in M$ an isometry. 
We now consider the case $\xi = v\Om$ and $\f = (\xi, \cdot \xi)$; that is, $\f(x) = \o(v^* x v)$, $x\in M$. 
\bprop\label{propv}
Let $M$, $\o$ and $\f = (\xi, \cdot \xi)$ be as above, where $\xi = v\Om$ with $v\in M$ an isometry.  Then
\ben\label{Sv}
S(\f |\!| \o) = -(\xi, \log\Delta_\Om\, \xi) \, .
\een
\eprop
\proof
We have $(\xi, x'\xi) =( v\Om,  x' v\Om) = (\Om, x'\Om)$, $x'\in M'$; that is, the vector states $\xi$ and $\Om$ agree on $M'$. Therefore, 
\[
\Delta(\f'/\o) = \Delta(\o'/\o) = \Delta_\Om\, ,
\] 
with $\o' = (\Om, \cdot \Om)$ on $M'$. So \eqref{Sv} follows by the definition \eqref{Edef},  
\eproof
We shall use the following known lemma.
\blem\label{v}
Let $M$ be a von Neumann algebra on the Hilbert space $\H$, with a cyclic and separating unit vector $\Om$.  With $\xi\in\H$ a unit vector, there exists an isometry $v\in M$ such that $\xi = v\Om$ iff the states $(\xi, \cdot \xi)$ and $(\Om, \cdot \Om)$ have the same restriction to $M'$. 
\elem
\proof
If $(\xi, \cdot \xi)|_{M'} = (\Om, \cdot \Om)|_{M'}$, the isometry $v$ is determined by $vx'\Om = x'\xi$, $x'\in M$. The other direction is clear. 
\eproof
We shall denote by
\[
S(\xi |\!| \Om)_M = S(\f|_{M} |\!| \o |_{M}) 
\]
the relative entropy of between the restrictions of $\f = (\xi, \cdot \xi)$ and $\o = (\Om,\cdot \Om)$ to a von Neumann algebra $M$ on $\H$. 

\subsection{Translation covariant nets of standard subspaces}
In the following,  a {\it wedge} $W$ of the Minkowski space $\RR^{1 +d}$ is a region obtained from $W_0 = \{x : x_1 > x_0\}$ by a spacetime translation, a space rotation and a space reflection. Thus, if $W$ is a wedge, also its spacelike complement $W'$ is a wedge.  
We denote by $\W$ the set of all wedges of $\RR^{1 +d}$.

A translation covariant {\it net of standard subspaces on wedges} on the Hilbert space $\H$ is a isotonous map
\[
H: W\in \W \mapsto H(W)\subset \H
\]
with $H(W)$ a standard subspace, together with a {\it positive energy unitary representation} $U$ of $\RR^{1 + d}$ such that
\[
U(x) H(W) = H(W +x)\, , \ x\in \RR^{1 + d}\, ,\ W\in \W\, ;
\]
the positive generator $P$ of the time-translation one-parameter unitary group is the {\it Hamiltonian} or {\it energy operator}. 

$H$ is {\it local} if $H(W') \subset H(W)'$, $W\in \W$. 

$H$ satisfies {\it wedge duality} if $H(W') = H(W)'$, $W\in \W$. 

A {\it net of real linear subspaces of $\H$} on $\H$ is defined similarly as above if theclosed, real linear subspace  $H(B)$ of $\H$ is defined also if the spacetime region $B\subset \RR^{1 +d}$ is Minkowski convex, that is $B$
the intersection of a family wedge regions. We assume $H(W)$ is standard for a wedge $W$, but we do not require $H(B)$ to be cyclic, if $B$ is bounded. (Examples, where $H(B)$ is not cyclic, are constructed in \cite{LL15} and \cite{LMR16}). 
If $B$ is any subset of $\RR^{1 +d}$, we denote by $H(B)$ the closed, real linear space generated by $H(C)$ as $C$ runs on the Minkowski convex regions $C\subset B$.
We set 
$\Delta_{B} = \Delta_{H(B)}$. 

With $R >0$, denote by $I_R$ the causal envelope of the spatial strip $\{ -R \leq x_1 \leq R\}$; thus $I_R$ is the intersection of two wedges with parallel edges: $I_R = W'_R\cap W_{-R}$, with $W_a = W_0 + (0, a, 0\dots 0)$. 

A {\it tube} of width $R$ is a spacetime translated of $I_R$. 
If a region $B\subset \RR^{1+d}$ is contained in a tube of radius $R$, and not in a smaller tube, we say that the {\it width} of $B$ is equal to $2R$. (Lorentz transformed tubes could be as well analysed by considering a suitable frame.)

Our result relies on the basic inequality \eqref{T2} due to \cite[Lemma 3.6]{BDL90} and \cite[Proposition 4.1]{BDL07} (see also \cite[Proposition 6.8]{L08}).   
We include a proof in Appendix \ref{M1P}. 
\bthm\label{M1}
Let $H$ be a local, translation covariant net of closed, real-linear subspaces  on $\H$, with Hamiltonian $P$, and $B$  a region of finite width $2R$. 
Then, 
\ben\label{T2}
||e^{-R\tan(2\pi \a)P}\xi|| \leq ||\Delta_B^{\a}\xi|| 
\een
for all $\xi\in \ov{H(B)}_\CC$ and  $0 < \a <1/4$. 
In particular, if $H(B)$ is standard, we have
\ben\label{T1}
e^{-2R\tan(2\pi \a)P} \leq  \Delta^{2\a}_{B}\, ,\quad 0 < \a <1/4\, .
\een
\ethm
\noindent
Note that neither duality nor the Bisognano-Wichmann property is assumed in the above theorem. 
\bcor\label{DPineq}
Let $H$ and $B$ be as in Theorem \ref{M1}, with  $H(B)$ standard. Then 
\ben\label{incor}
-\log \Delta_B \leq 2\pi \text{\small R}\, P\, .
\een
\ecor
\proof
By Proposition  \ref{A<B} and eq. \eqref{T1}
we have
\[
-(\xi , \log \Delta_B \xi ) \leq  R\frac{\tan(2\pi \a)}{\a} \, (\xi ,P\xi )\, .
\]
Letting $\a \to 0^+$, we get \eqref{incor}. 

The general case follow by \eqref{T2}; if $\xi\in H(B)_\CC$,  
\[
(\xi , \log \Delta_B \xi )  = \lim_{\a\to 0^+} \big( ||\Delta_B^{\a}\xi||^2 - 1\big)/2\a \geq \lim_{t\to 0^+} \big( ||e^{-R\tan(\pi \a)P}\xi|| ^2 - 1\big)/2\a
= 2\pi  \textit{\small R}\, (\xi ,P\xi )\, ,
\]
by \eqref{At}. 
\eproof
\section{Relative entropy/energy ratio bound}
Let $\H$ be a Hilbert space and $\A$ {\it a net of von Neumann algebras} on $\H$ on the Minkowski spacetime $\RR^{1 + d}$.  Namely,
$\A$ is a isotonous map
\[
O\mapsto \A(O)
\]
from the family $\O$ of double cones of   $\RR^{1 + d}$ to the family of von Neumann algebras on $\H$. 

If $B\in  \RR^{1 + d}$ is any region, $\A(B)$ is then defined as the von Neumann algebra generated by all $\A(O)$'s with $O\subset B$, $O\in \O$. 

$\A$ is {\it translation covariant} if there exists a {\it positive energy} representation $U$ of  $\RR^{1 + d}$ on $\H$ such that
$U(x)\A(O)U(-x) = \A(O + x)$, $O\in\O$, $x\in \RR^{1 + d}$; and there is a $U$-invariant unit vector $\Om\in\H$, the {\it vacuum vector}, with $\Om$ is cyclic for $\A(W)$ if $W$ is a wedge. 

$\A$ is {\it wedge dual} if $\A(W') = \A(W)'$ if $W$ is a wedge.\footnote{Wedge duality holds automatically in a Wightman theory \cite{BW75}.}

If $\A$ is wedge dual, and $B\subset\RR^{1 + d}$ is a region, we set 
\[
\A^d(B) = \A(B')' \, .
\]
Then $O\in\O \mapsto \A^d(O)$ is a net on double cones, {\it the dual net} of $\A$. The net $\A^d$ is local,  $\A(O) \subset \A^d(O)$,
$O\in\O$,  and $\A^d(W) = \A(W)$, with $W$ a wedge. 

If $B$ is Minkowski convex, so $B$ is the intersection of a family of wedge regions $W_i$, then $B' = \cup_i W'_i$, so
\[
\A^d(B) = \bigcap_i \A(W_i)\
\]
by wedge duality; Haag duality for $\A^d$ holds for Minkowski convex regions $B$:
\[
\A^d(B') =  \A^d(B')\, .
\]
Let $\xi\in\H$ be a unit vector and $\f = \langle\cdot\rangle_\xi$ the associated pure state of $B(\H)$. We shall say that the vector $\xi$ is {\it localized in the region $B$} if $\f= \o$ on $\A(B')$, with $\o = \langle\cdot\rangle_\Om$ the vacuum state. 
\blem\label{lemin}
If $\A$ is wedge dual and $B$ is Minkowski convex, the unit vector $\xi\in\H$ is localized in $B$ iff there exists an isometry $v\in \A^d(B)$ such that $\xi = v\Om$. 
\elem
\proof
Immediate by Lemma \ref{v} and $\A^d(B)' = \A(B')$. 
\eproof
Lemma \ref{lemin}, and the entropy bound in \cite{LM24}, will allow us to pass to the dual net. 
Our main result is the following theorem. 
\bthm\label{main}
Let $\A$ be a wedge dual, translation covariant net of von Neumann algebras,
and $B$ be a spacetime region of width $2R$.  
If $\xi$ is a unit vector localized in $B$, then
\[
S(\xi |\!| \Om)_{\A(B)}   \leq 2\pi R\,  \langle P\rangle_\xi  \, ,
\]
with $P$ the Hamiltonian and $\langle P\rangle_\xi = (\xi, P\xi)$. 
\ethm 
\proof
By translation covariance, we may assume that $B$ is contained in the tube $I_R$. 

We consider the net of closed, real linear subspaces $H(B) = \ov{\A^d(B)_{\rm s.a.}\Om}$; note that
\ben\label{AH}
\Delta_{H(B)} = \Delta_{\A^d(B)}\, .
\een
Then,
\begin{align*}
S(\xi |\!| \Om)_{\A(B)}  &\leq S(\xi |\!| \Om)_{\A(I_R)} &\text{(monotonicity of relative entropy)}\\
 &\leq S(\xi |\!| \Om)_{\A^d(I_R)} &\text{(monotonicity of relative entropy)}\\
  &= -( \xi, \log \Delta_{\A^d(I_R)} \xi) &\text{(Lemma \ref{lemin} and Proposition \ref{propv})}\\
   &= -( \xi, \log \Delta_{H(I_R)} \xi) &\text{(identity \eqref{AH})}\\
 &\leq 2\pi R\, \langle P\rangle_\xi  &\text{(Corollary \ref{DPineq})}
\end{align*} 
so we have our inequality. 
\eproof
With $\f = \langle \cdot\rangle_\xi$, we define the {\it energy of the state $\f$ in the region $B$} by
\[
E(\f; B) =  \inf \big\{\langle P\rangle_\eta : \eta\in\H, \,\langle\cdot\rangle_\eta |_{\A(B)} 
= \f\big\} \, .
\]
\bcor
In the above setting, we have
\[
S(\f |\!| \o)_{\A(B)}   \leq 2\pi \textit{\small R} \, E(\f; B)\, .
\]
\ecor
\proof
Immediate from Theorem \ref{main}, since 
$S(\xi |\!| \Om)_{\A(B)}$ depends only on the restriction $\f$ of the state $\langle\cdot\rangle_\xi$ to the local algebra $\A(B)$. 
\eproof

\section{Final comments}
Since our bound essentially follows from locality and translation covariance with positivity energy, one may wonder whether it holds in a classical context too. The quantum structure in this paper is manifest by the non-commutativity of the local algebras; if the local algebras were commutative, then the translation symmetries would be trivial \cite{L82}. Said differently, the vacuum Reeh-Schlieder cyclicity property evinces the quantum fluctuations and guarantees a non trivial structure. 

However, our bound applies to the classical Klein-Gordon or Dirac field theory too. One of our starting points was indeed to estimate the massive modular Hamiltonian of a free quantum particle. 
We have written part of this paper from the more general standard subspace viewpoint (a direct analysis with local nets of von Neumann algebras would have been possible) both to continue our analysis in this direction, and because this setting naturally arires in other contexts and is not more elaborate. 

Note also that, although Poincaré covariance is not needed for our result, a two-dimen\-sional Poincaré group emerges from modular theory by Borchers’ theorem. 

We plan to extend our analysis concerning Fermi nets and non globally pure states, a first study is contained in \cite{L20}. We are also going to provide a form of our bound for approximately localized states, taking into account quantum fluctuations. We will get to these and other points somewhere else.

\appendix
\section{Appendix}\label{M1P}
For the reader's convenience, we provide a proof of Theorem \ref{M1}. 

With $\H$ a Hilbert space and $H$ be a two-dimensional net of standard subspaces of $\H$ on wedges, we shall set $H(a, \infty) = H(W_a)$, $H(- \infty, a) = H(W'_a)$, $a\in \RR$, where $W$ is the right wedge $W_a = W_0 + (0,a)$. We simplify our notations for the modular operators by writing $\Delta_a = \Delta_{H(W_a)}$, $\Delta'_a = \Delta_{H(W'_a)}$. 
\bprop
Let $H$ be a 2-dimensional, translation covariant, net of standard subspaces on $\H$, with $U$ the spacetime translation unitary group. With $R >0$ and $P$ the generator of the time translation one-parameter unitary group,
we have  
\begin{align}
||E_+ e^{-R\tan(2\pi \a)P}\xi || & \leq  || \Delta^{\a}_{-R}\xi||\, , \label{Pnorm01b}
\\
||E_- e^{-R\tan(2\pi \a)P}\xi ||  & \leq || \Delta'^{\a}_{R}\xi|| 
\, ,\label{Pnorm02b}
\end{align}
$0< \a < 1/4$, for all $\xi\in \H$,
with $E_\mp$ the spectral projections of $\Delta_{(0,\infty)}$ associated with $[0, 1)$ and  $[1,\infty)$ 
\eprop
\proof
Let $0 < a < R$. 
By Theorem \ref{BH} we have
\ben\label{DU}
\Delta^{is}_{-a}  = U(0, -a) \Delta^{is}_0 U(0, a)  
= U(0, -a)U\big(\L(-2 \pi s)(0,a)\big)  \Delta^{is}_0   \, ,
\een
where $\L(s) = \begin{pmatrix} \cosh s & \sinh s\\ \sinh s & \cosh s \end{pmatrix}$, $s\in \RR$. 
Therefore, by \eqref{DU}, we have
\ben\label{sz}
T_{(-a,\infty),(0,\infty)}(is) =  
\Delta^{is}_{-a}  \Delta^{-is}_0  
= U\big((-a\sinh 2\pi s, a\cosh 2\pi s -a)\big) 
   \, , \quad s\in \RR
\een
By Lemma \ref{T}, the map $s\in\RR \mapsto T_{(-a,\infty),(0,\infty)}(is)$ is the boundary value of a holomorphic family of contractions $T_{(-a,\infty),(0,\infty)}(iz)$ on the strip  $-1/2 < \Im z < 0$, continuous on the closure of the strip. By the positivity of the energy, $U$ is the boundary value of an analytic operator values function on $\RR^2 - i V_+$, with $V_+$ the forward light cone. 

By analytic continuation, setting $s = -i\a$, $0 < \a < 1/4$ in \eqref{sz}, we have
\[
\ov{\Delta^{\a}_{-a}  \Delta^{-\a}_0}
\subset U\big((-ia\sin 2\pi \a, a\cos 2\pi \a -a)\big) 
= U(0,R -a)e^{- R\tan(2\pi\a) P}\, ,
\]
where, in the last term, we have chosen $a  = R/\cos 2\pi\a$;
thus, by multiplying both sides of the above equality on the left by $U(0,a -R)$,
\[
\Delta^{\a}_{-R} U(0,a -R) \Delta^{-\a}_0  \subset e^{-R\tan(2\pi \a)P} \, ;
\]
taking adjoints and using \eqref{TC} we then get
\[
 \Delta^{-\a}_0  U(0,R - a)  \Delta^{\a}_{-R} \subset e^{-R\tan(2\pi \a)P}\, ,
\]
that is
\[
 \Delta^{-\a}_0 U(0,R - a)  \Delta^{\a}_{-R}\xi  = e^{-R\tan(2\pi \a)P}\xi
\]
if $\xi\in D(\Delta^{\a}_{(-R,\infty)})$. This implies that 
\[
||E_+ e^{-R\tan(2\pi \a)P}\xi|| = ||E_+  \Delta^{-\a}_0 U(R - a)  \Delta^{\a}_{-R}\xi ||\leq || \Delta^{\a}_{-R} \xi ||
\]
for $\xi\in D(\Delta^{\a}_{-R})$, hence for all $\xi\in \H$. 

The inequality \eqref{Pnorm02b} follows similarly by considering the left wedge $W'_{(0,R)}$ instead of $W_{(0,-R)}$. 
\eproof
Let now $H$ be a two-dimensional translation covariant net of standard subspace on $\H$, and denote by
 $H(-R,R)$  the closed, real linear subspace associated with the double cone $O_R$ centered at $(0, 0)$ with radius $R$, so $O_{R} = W_{-R}\cap W'_{R}$ and set $\Delta_{(-R,R)} = \Delta_{H(O_R)}$. 
 \bcor
 Let $H$ be a two-dimensional, translation covariant net of standard subspaces on $\H$. If $H$ is local, we have
 \ben\label{ineqf}
||e^{-R\tan(2\pi \a)P} \xi || \leq ||\Delta^{\a}_{(-R, R)}\xi|| \, ,\quad \xi\in \ov{H(-R,R)}_\CC\, . 
\een
The inequality \eqref{ineqf} holds also if $H$ is not local if $\Delta'_0 = \Delta^{-1}_0$. 
 \ecor
 \proof
If $H$ is local, by re-defining the left lines spaces by $\bar H(-\infty, a) = H(a,\infty)'$, we may assume that duality for half-lines holds. 
Then $\Delta'_0 = \Delta^{-1}_0$, so we assume this relation to be satisfied.  

Then, by \eqref{Pnorm01b}, we have
\[
||E_+ e^{-R\tan(2\pi \a)P}\xi || \leq  
|| \Delta^{\a}_{-R}\xi|| = 
 || T_{(-R,\infty), (-R,R)}(\a) \Delta^{\a}_{(-R,R)}\xi ||
 \leq ||\Delta^{\a}_{(-R,R)}\xi || \, ;
\]
and similarly, by   \eqref{Pnorm02b}, we have
\[
||E_- e^{-R\tan(2\pi \a)P}\xi || \leq ||\Delta^{\a}_{(-R,R)}\xi ||\, ,
\]
and $E_-$ is the spectral projection of $\Delta_0$ associated with $(0,1]$. 

By putting together these inequalities, we get
\ben\label{ineqA}
|| e^{-R\tan(2\pi \a)P}\xi || \leq 2||\Delta^{\a}_{(-R,R)}\xi ||\, ,
\een
for all $\xi\in \H$. 

Now, the inequality \eqref{ineqA} also holds for the net $H\otimes H \otimes\cdots \otimes H$ of $\H\otimes \H \otimes\cdots \otimes \H$ and the vector $\xi\otimes \xi \otimes\cdots \otimes \xi$
($n$ copies), thus
\[
||e^{-R\tan(2\pi \a)P}\xi ||^n \leq 2 || \Delta^{\a}_{(-R, R)}\xi||^n
\]
so \eqref{ineqA} can be sharpened to
\[
||e^{-R\tan(2\pi \a)P} \xi || \leq ||\Delta^{\a}_{(-R, R)}\xi|| \, ,
\]
for all $\xi\in \ov{H(-R,R)}_\CC$.  
 \eproof
\smallskip

\noindent{\bf Proof of Theorem \ref{M1}.} By changing the spacetime frame, if necessary, we may assume that $B$ is contained in the tube $I_R$. 

We consider the two-dimensional net $K$ obtained by projecting the net $H$ onto the $x_0\, x_1$-plane. Namely, if $C$ is a region of the $x_0\, x_1$-plane, we put
\[
K(C) = H(\tilde C)\, ,
\]
where $x\in \tilde C$ iff $(x_0, x_1)\in C$.  

By \eqref{ineqf}, we have
\[
||e^{-R\tan(2\pi \a)P} \xi || \leq ||\Delta^{\a}_{K(-R, R)}\xi|| = ||\Delta^{\a}_{H(I_R)}\xi|| \, ,
\]
for all $\xi\in \H$. 
As $B\subset I_R$, we then have
\[
||e^{-R\tan(2\pi \a)P} \xi||  = ||T_{I_R, B}(\a)\Delta^{\a}_{B}\xi|| \leq  ||\Delta^{\a}_{B}\xi|| \, ;
\]
that is, $e^{-2R\tan(2\pi \a)P} \leq  \Delta^{2\a}_{B}$, if $0 \leq \a \leq1/4$, as desired. 
\eproof

\bigskip

\noindent
{\bf Acknowledgements.} 
We thanks R. Bousso, H. Casini and S. Hollands for valuable comments. 

\noindent
We acknowledge the MIUR Excellence Department Project awarded to the Department of Mathematics, University of Rome Tor Vergata, CUP E83C18000100006, and support of the Institut Henri Poincaré (UAR 839 CNRS-Sorbonne Université), and LabEx CARMIN (ANR-10-LABX-59-01). 


\begin{thebibliography}{99}\itemsep-2pt

\bibitem{Ar76} {\sc H. Araki},
{\it Relative entropy of states of von Neumann algebras},
Publ. RIMS Kyoto Univ. 11 (1976), 809--833. 

\bibitem{Ar77} {\sc H. Araki},
{\it Relative entropy of states of von Neumann algebras. II},
Publ. RIMS Kyoto Univ. 13 (1977), 173--192. 

\bibitem{B81} {\sc J.D. Bekenstein},
{\it Universal upper bound on the entropy-to-energy ratio for bounded systems},
Phys. Rev. D23 (1981) 287.

\bibitem{BW75}  {\sc J. Bisognano, E. Wichmann}, 
{\it On the duality condition for a Hermitean scalar field}, 
J. Math. Phys. 16 (1975), 985.

\bibitem{BC13}  {\sc D.D. Blanco, H. Casini},
{\it Localization of negative energy and the Bekenstein bound},
Phys. Rev. Lett. 111 (2013) 221601.

\bibitem{Bo92}  {\sc H.J. Borchers}, 
{\it The CPT theorem in two-dimensional theories of local observables},
 Comm. Math. Phys. 143, 315--332 (1992)

\bibitem{B03} {\sc R. Bousso},
{\it Light-sheets and Bekenstein’s bound},
Phys. Rev. Lett. 90, 121302 (2003). 

\bibitem{BDL190} {\sc D. Buchholz, C. D’Antoni, R.  Longo},
{\it  Nuclear maps and modular structures I. General properties},
J.  Funct.  Anal.  88 (1990), 223-250.

\bibitem{BDL90} {\sc D. Buchholz, C. D’Antoni, R.  Longo}, 
{\it Nuclear maps and modular structures. II Applications to quantum field theory},
Comm. Math. Phys. 129 (1990), 115--138.

\bibitem{BDL07} {\sc D. Buchholz, C. D’Antoni, R.  Longo}, 
{\it Nuclearity and thermal states in Conformal Field Theory}, 
Comm. Math. Phys. 270 (2007), 267--293.

\bibitem{Ca08} {\sc H. Casini},
{\it Relative entropy and the Bekenstein bound},
Class. Quantum Grav. 25 (2008), 205021. 

\bibitem{H} {\sc R. Haag}, 
``Local Quantum Physics -- Fields, Particles, Algebras'',   
2nd edn., Springer, New York (1996).

\bibitem{HW} {\sc  P. Hayden, J. Wang},
{\it What exactly does Bekenstein bound?},
arXiv:2309.07436

\bibitem{LL15} {\sc G. Lechner, R.  Longo},
{\it  Localization in nets of standard spaces},
Comm. Math. Phys. 336 (2015), 27--61.

\bibitem{L82}{\sc R. Longo}, 
{\it Algebraic and modular structure of von Neumann algebras of Physics}, 
Proc. Symp. Pure Math. 38 (1982), Part 2, 551.

\bibitem{L97}{\sc R. Longo}, 
{\it An analogue of the Kac-Wakimoto formula and black hole conditional entropy}, 
Comm. Math. Phys. 186 (1997), 451--479.

\bibitem{L08}  
{\sc R. Longo}, 
{\it Real Hilbert subspaces, modular theory, $SL(2, R)$ and CFT}, 
in: ``Von Neumann algebras in Sibiu'',  Theta Ser. Adv. Math., 10, 33--91, Theta, Bucharest, (2008).

\bibitem{L18} 
{\sc R. Longo}, 
{\it On Landauer’s principle and bound for infinite systems},
Comm. Math. Phys. 363 (2018), 531--560. 

\bibitem{L20} 
{\sc R. Longo}, 
{\it Entropy distribution of localised states}, 
Comm. Math. Phys. 373, 473--505 (2020).  

\bibitem{LM24} {\sc R. Longo, V. Morinelli}, 
{\it An entropy bound due to symmetries},
Rev. Math. Phys. Special Issue in Memory of H. Araki (in press). 

\bibitem{LMR16} {\sc R. Longo, V. Morinelli, K.H. Rehren}, 
{\it Where infinite spin particles are localized}, 
Comm. Math. Phys. 345 (2016), 587–614.

\bibitem{LX18} {\sc R. Longo, F. Xu}, 
{\it Comment on the Bekenstein bound}, 
J.  Geom.  Phys. 130, 113--120, (2018).

\bibitem{OP}{\sc M.  Ohya, D. Petz},
``Quantum entropy and its use", 
Texts and Monographs in Physics. Springer-Verlag, Berlin, 1993,

\bibitem{P08}{\sc D.N. Page},
{\it Defining entropy bounds},
JHEP 10 (2008), 007. 

\bibitem{Sch}{\sc K. Schmüdgen}
``Unbounded Self-adjoint Operators on Hilbert Space",
Springer-Verlag, New York-Heidelberg, 2012.

\bibitem{Str} {\sc S. Stratila},
``Modular theory in operator algebras",
Cambridge Univ. Press 2021. 

\bibitem{Tak} {\sc M. Takesaki}, 
``Theory of operator algebras'', I \& II, 
Springer-Verlag, New York-Heidelberg, 2002 \& 2003.
\end{thebibliography}
\end{document}